\documentclass[preprint,pra,showpacs,preprintnumbers,amsmath,amssymb]{revtex4}

\usepackage{graphicx}
\usepackage{dcolumn}
\usepackage{bm}
\usepackage{subfig}

\begin{document}

\def\kket{\rangle \mskip -3mu \rangle}
\def\bbra{\langle \mskip -3mu \langle}

\def\ket{\rangle}
\def\bra{\langle}

\def\pard{\partial}

\def\sinh{{\rm sinh}}
\def\sgn{{\rm sgn}}

\def\alp{\alpha}
\def\del{\delta}
\def\Del{\Delta}
\def\eps{\epsilon}
\def\gam{\gamma}
\def\sig{\sigma}
\def\kap{\kappa}
\def\lam{\lambda}
\def\ome{\omega}
\def\Ome{\Omega}

\def\vphi{\varphi}

\def\Gam{\Gamma}
\def\Ome{\Omega}

\def\kav{{\bar k}}
\def\vb{{\bar v}}

\def\abf{{\bf a}}
\def\cbf{{\bf c}}
\def\dbf{{\bf d}}
\def\gbf{{\bf g}}
\def\kbf{{\bf k}}
\def\lbf{{\bf l}}
\def\nbf{{\bf n}}
\def\pbf{{\bf p}}
\def\qbf{{\bf q}}
\def\rbf{{\bf r}}
\def\ubf{{\bf u}}
\def\vbf{{\bf v}}
\def\xbf{{\bf x}}
\def\Cbf{{\bf C}}
\def\Dbf{{\bf D}}
\def\Kbf{{\bf K}}
\def\Pbf{{\bf P}}
\def\Qbf{{\bf Q}}

\def\omet{{\tilde \ome}}
\def\gammat{{\tilde \gamma}}
\def\Ft{{\tilde F}}
\def\gt{{\tilde g}}
\def\Ht{{\tilde H}}
\def\ttil{{\tilde t}}
\def\Ut{{\tilde U}}
\def\ut{{\tilde u}}
\def\bt{{\tilde b}}
\def\Vt{{\tilde V}}
\def\vt{{\tilde v}}
\def\xt{{\tilde x}}

\def\ph{{\hat p}}

\def\vt{{\tilde v}}
\def\wt{{\tilde w}}
\def\phit{{\tilde \phi}}
\def\rhot{{\tilde \rho}}
\def\Ft{ {\tilde F}}

\def\Cb{{\bar C}}
\def\Nb{{\bar N}}
\def\Ab{{\bar A}}
\def\Db{{\bar D}}
\def\etab{{\bar \eta}}
\def\gb{{\bar g}}
\def\nb{{\bar n}}
\def\bb{{\bar b}}
\def\Pib{{\bar \Pi}}
\def\rhob{{\bar \rho}}?\def\phib{{\bar \phi}}
\def\psib{{\bar \psi}}
\def\omeb{{\bar \ome}}

\def\Sh{{\hat S}}
\def\Wh{{\hat W}}

\def\SS{I}
\def\psiw{{\xi}}
\def\tI{{g}}

\def\Ep#1{Eq.\ (\ref{#1})}
\def\Eqs#1{Eqs.\ (\ref{#1})}
\def\EQN#1{\label{#1}}

\newcommand{\beqa}{\begin{eqnarray}}
\newcommand{\eeqa}{\end{eqnarray}}

\title{Decay modes of two repulsively interacting bosons}

\author{Sungyun Kim$^1$}
 \email{rdecay@googlemail.com}
\address{$^1$ Hoseo University, 165 Sechul Li, Baebang Myun, Asan, Chungnam 336-795, Korea}
\author{Joachim Brand$^2$}
\email{J.Brand@massey.ac.nz}
\address{$^2$ Centre for Theoretical Chemistry and Physics and
New Zealand Institute for Advanced Study,
Massey University,
Private Bag 102904, North Shore,
Auckland 0745, New Zealand}

\begin{abstract}
We study the decay of two repulsively interacting bosons tunneling through a delta potential barrier by direct numerical solution of the time-dependent Schr\"odinger equation.
The solutions are analyzed according to the regions of particle presence:
both particles inside the trap (in-in), one particle in and one particle out (in-out),
 and both particles outside (out-out). It is shown that the in-in probability is dominated by exponential decay, and its decay rate is predicted very well from outgoing boundary conditions.
 Up to a certain range of interaction strength the decay of in-out probability is dominated
 by the single particle decay mode.
 The decay mechanisms are adequately described by simple models.

\end{abstract}

\maketitle

\section{Introduction}
 The decay of a particle by tunneling through a potential barrier into a continuum is a fundamental and unique phenomenon in quantum mechanics. The tunneling of multi-particle systems is just as important and presents one of the places where the understanding of macroscopic
quantum  phenomena can start \cite{BPAnderson}.
The tunneling and decay of Bose-Einstein condensates (BECs) are attractive subjects of study \cite{Leggett},
 since the BEC is a unique state of matter where quantum mechanical features are manifested at the macroscopic level.
 After BECs were first realized experimentally in dilute atomic gases \cite{BEC1st}, a huge amount of related research followed. Ultra-cold atoms are usually trapped in a finite potential well and the decay by tunneling into a continuum is an existing and potentially desirable possibility. In this context it was realized that  understanding the decay dynamics by tunneling is a very important task  \cite{Raizen1, Raizen2}.

In most cases BECs have thousands to millions of particles and the dynamics is adequately described by the Gross-Pitaevskii (GP) equation \cite{Gross,Pitaevki}  of mean-field theory \cite{Pethick},
a nonlinear Schr\"{o}dinger equation.
The GP equation governs the time evolution of phase and particle number density of an essentially fully Bose-condensed system. With many works on the mean-field description of BEC tunneling \cite{Moiseyev04, Moiseyev05,Fleurov,Schlagheck,Carr05}, it is remarkable that there is still a discussion, both about the technical implementation \cite{Schlagheck} and  the correct formulation of mean-field theory related to the decay problem \cite{Moiseyev05}. It is thus desirable to obtain a detailed understanding of the microscopic physics of multi-particle decay.


  The cases of stronger interactions or fewer particle numbers are also important,
 where the GP equation is less accurate. In the few boson regime, correlated decay of particles was
 observed and studied both experimentally and theoretically \cite{Foelling, Zoellner}. The particle correlation
 in the decayed wave is important in relation to the atom laser \cite{Oettl}.
 For strongly-interacting bosons in a one-dimensional trap Bose-condensation is not relevant but the gas acquires properties related to fermionic systems \cite{Girardeau}. In the Tonks-Girardeau limit of infinite interactions the few boson decay problem was treated analytically \cite{Campo},
and numerical simulation have addressed the crossover for finite interactions from a harmonic trap with up to four bosons \cite{Lode}. The analytic treatment of
 few boson decay with finite interaction strength
 remains a difficult task.

 In this paper, we approach this problem by both numerically and analytically.
 We study the simplest case of two repulsively interacting bosons in a potential
trap in one dimension. The time evolution of the decay is obtained from first principles by solving
the time-dependent Schr\"{o}dinger equation numerically. Then, it is compared to approximate analytic methods,
starting from the exact solutions of local spatial regions. The decay phenomena are investigated for a wide range
of interaction strength, from zero to very strong repulsion.
The analytic model predicts
exponential decay mode of the interacting system, which is in very good agreement with our numerical simulation. Also, the decay of the total particle number
is well explained with a simple theoretical model.

\section{The model Hamiltonian}
  \label{apdx:1}

 We choose a model Hamiltonian for the two interacting boson
 decay. Considering the kinetic energy, external potential $\Vt_{ex}$ for trapping and interaction $\Ut$ between bosons,
  the total Hamiltonian with two identical bosons is written as
    \beqa
    \Ht = -\frac{\hbar^2}{2m}
 \frac{\partial^2}{\partial \xt_1^2}  -\frac{\hbar^2}{2m}
 \frac{\partial^2}{\partial \xt_2^2} + \Vt_{ex}(\xt_1) + \Vt_{ex}(\xt_2) +
    \Ut(\xt_1,\xt_2).
    \eeqa
 The external potential is
 \beqa
 \Vt_{ex} (\xt) =\left\{ \begin{array}{ll} \infty
   \;& \mbox{for $\xt<0$} \\
   \Vt \delta (\xt-L) \;& \mbox{for $\xt \ge 0$} \end{array} \right.
 \eeqa
 and acts as potential trap by a delta barrier at position $L$. This choice of external potential has some advantages in that the delta barrier has zero width so the consideration of decay process inside the barrier is not needed. Also, the analytical treatment of the decay process is simplified. In single particle case we found that the exponential decay mode dominates and non-exponential features are strongly suppressed compared to a finite-width barrier case. Computationally, the narrow width of the delta function  makes the Hamiltonian matrix more sparse, which makes the problem  tractable.

  Considering only s-wave scattering \cite{Leggett}, the interaction
 potential between particle 1 and particle 2 is given as
  \beqa
   \Ut (\xt_1,\xt_2) = \gt \,\delta(\xt_1-\xt_2)
   \eeqa
 where $\gt$ is a coupling constant and $\xt_1$ and $\xt_2$ are the
 positions of each boson, respectively.

  To simplify the analysis and compare the result with external parameters, we introduce dimensionless units. The new length unit $x$
is defined as $x \equiv \xt/L$. The Hamiltonian is rewritten as
\beqa
& &\Ht= -\frac{\hbar^2 }{2 m L^2} \frac{\partial^2}{\partial x_1^2}   -\frac{\hbar^2 }{2 m L^2} \frac{\partial^2}{\partial x_2^2}
+ \frac{\Vt }{L} \delta (x_1-1) +  \frac{\Vt }{L} \delta (x_2-1)  \nonumber \\
& &+ \frac{\gt}{L} \delta (x_1-x_2).
\;\;\; \mbox{for $x \ge 0$}
\eeqa
Dividing both sides by $\hbar^2 / (m L^2)$, we get the rescaled, dimensionless Hamiltonian $H \equiv  m L^2 \Ht / \hbar^2 $  .
 \beqa
 H = -\frac{1}{2} \frac{\partial^2}{\partial x_1^2}   -\frac{1}{2} \frac{\partial^2}{\partial x_2^2}
 + V \delta (x_1-1) + V \delta (x_2-1) + g \delta (x_1 - x_2)
 \eeqa
 Here
 \beqa
 V \equiv \frac{m L}{\hbar^2 } \Vt, \;\;\;
 g \equiv \frac{m L}{\hbar^2 } \gt.
 \eeqa
 The Schr\"{o}dinger equation with this Hamiltonian is given by
 \beqa
 i\partial_t \psi = H \psi,
 \eeqa
 where $t= \hbar \,\ttil/ (m L^2)$ with $\ttil$ is unscaled time.

\begin{figure}[hbp]
\includegraphics[scale=.7]{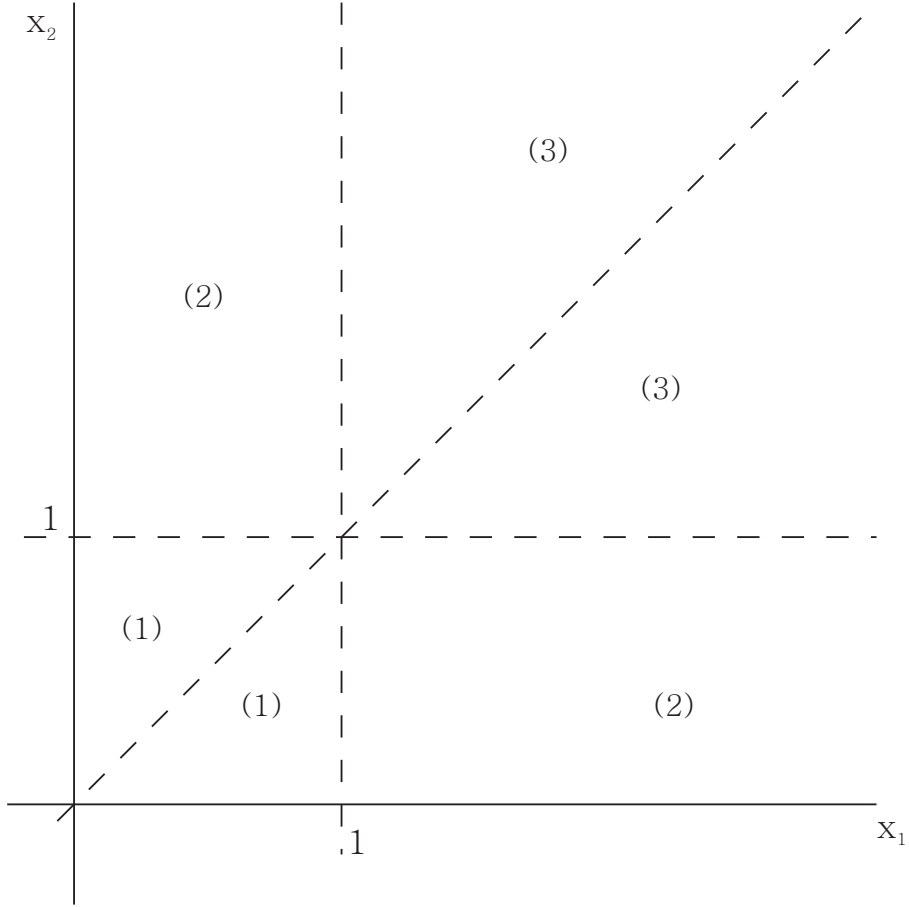}
\caption{Hamiltonian in $x_1-x_2$ space. }
\label{setup2}
\end{figure}

 In $x_1-x_2$ space, the Hamiltonian looks like figure~\ref{setup2}. The dotted lines represent delta potentials from trap and interaction between particles. From now on we denote the region where both particles inside the trap as region (1), where one particle in and one particle out of the trap as region (2), and both particles out of trap as region (3).

\section{Numerical simulation of two boson decay}
\label{numerical}
\begin{figure}[hbp]
\includegraphics[scale=.7]{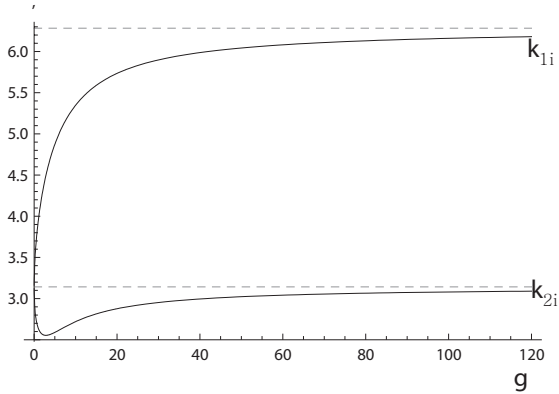}
\caption{$k_{1i}$ and $k_{2i}$ versus $g$ plot. The dashed lines are $\pi$ and $2 \pi$, the wavevectors of
$V=\infty$ lowest and next lowest states. As the interaction strength $g$ increases, $k_{1i}$ approaches to $2 \pi$ and
$k_{2i}$ approaches to $\pi$.  }
\label{kgplot}
\end{figure}

 Now, we set up the decay of two interacting identical boson in this Hamiltonian. We choose the initial condition that both particles inside the delta trap as the two boson ground state of $V \rightarrow \infty$ case. Specifically, this initial state $\psi_{\rm ini} (x_1,x_2)$ is given by \cite{Gaudin}
  \beqa
 \psi_{\rm ini}(x_1,x_2)& &= N_{\rm ini} \big( (A_1(k_{1i},k_{2i}) e^{i k_{1i} x_1} -A_2(k_{1i},k_{2i}) e^{-i k_{1i} x_1}) \sin (k_{2i} x_2) \nonumber \\
 & &+ (A_3(k_{1i},k_{2i}) e^{i k_{2i} x_1} -A_4(k_{1i},k_{2i}) e^{-i k_{2i} x_1}) \sin (k_{1i} x_2) \big) \nonumber \\
   & &\mbox{ for $0 \le x_2 \le x_1 \le 1$}
  \eeqa
 with
 \beqa
  A_1(k_{1},k_{2})  = (i k_{1} + i k_{2} +g)(i k_{1} -i k_{2} +g) \EQN{A1} \\
  A_2(k_{1},k_{2}) = (i k_{1} -i k_{2} -g)(i k_{1} + i k_{2} -g)  \EQN{A2} \\
  A_3(k_{1},k_{2}) = (i k_{1} +i k_{2} +g)(i k_{1} -i k_{2} -g)  \EQN{A3}\\
  A_4(k_{1},k_{2}) = (i k_{1} +i k_{2} -g)(i k_{1} - i k_{2} +g). \EQN{A4}
  \eeqa
  $k_{1i}$ and $k_{2i}$ satisfy the equation
  \beqa
   & &k_{1i} = \pi + \arctan (\frac{g}{k_{1i}-k_{2i}} ) + \arctan( \frac{g}{k_{1i}+k_{2i}}), \\
   & &k_{2i}= \pi -\arctan (\frac{g}{k_{1i}-k_{2i} }) +\arctan (\frac{g}{k_{1i}+k_{2i}}).
  \eeqa
  The initial wave function in $0 \le x_1 \le x_2 \le 1$ region is obtained from the
 boson symmetry condition $\psi_{\rm ini} (x_1,x_2) = \psi_{\rm ini} (x_2,x_1)$. In other regions the initial wave function is zero.
 The normalization constant $N_{\rm ini}$ is chosen to satisfy $\int dx_1 dx_2 |\psi_{\rm ini}|^2 =1$.
 $k_{1i}$ and $k_{2i}$ versus interaction strength $g$ is shown in figure~\ref{kgplot}.
  For zero interaction both $k_{1i}$ and $k_{2i}$ are same as $\pi$, the single particle ground state
  wavevector. For nonzero $g$ they rapidly deviate form $\pi$ as $g$ increases, and $k_{1i}$ approaches to $2 \pi$ and $k_{2i}$
  approaches to $\pi$ (Figure~\ref{kgplot}) .

  To analyze the decay of interacting bosons, we solve the Schr\"{o}dinger equation directly.
  The Schr\"{o}dinger equation and its formal solution are
 \beqa
  & &i \partial_t \psi = H \psi \\
  & & \psi (t) = \exp (-iHt)\psi (0).
  \eeqa
    We use Crank-Nicolson method to solve this equation numerically \cite{CN1,CN2}.

  For the numerical representation of Hamiltonian, we choose the triangular region $0 \le x_2 \le x_1 \le X_{max}$ in $x$ space
  ($0 \le x_1 \le x_2 \le X_{max}$ region is determined due to the bosonic symmetry), with $X_{max}$ is large enough that in our observing time very little decay products
   reach near $X_{max}$.
This region is discretized by dividing $X_{max}$ by $N_x$, and
all points in the triangular region are arranged in one column vector.
 The Hamiltonian
  matrix obtained by discretization of $x$ space and using a finite-difference formula for the second derivatives can be quite large,
  but it is a sparse matrix as most elements are zero.

   For small $dt$,
  \beqa
    & &\exp (iH dt/2) \,\psi (t+dt) =  \exp (-iH dt/2)\, \psi(t) \\
    & &\psi (t+dt) = (1 +iH dt/2)^{-1} (1- iH dt/2)\, \psi (t) + O(dt^3). \EQN{CNmethod}
  \eeqa
   This method is second order in $dt$ and unitary (i.e.\ probability is conserved). This is an implicit method, since it contains the    inverse operator. The matrix inversion is efficiently implemented by solving linear equations. The time evolution of the wave function is obtained by iterating equation (\ref{CNmethod}).

   For the simulations in next sections, the following parameters are used. $X_{\rm max} =45$,
   $\Delta x=(X_{\rm max}/N_x) =  1/24$, $dt = 0.002$, $V =5$.
   The convergence of the numerical solutions is checked by changing spatial grid size and time step. We also check numerical simulation with known analytic solutions for special cases $g=0$ and $g=\infty$. To see the effect
    from the reflection of waves at the boundary the results are examined by changing $X_{max}$ and
    by putting absorbing potentials near $X_{\rm max}$. In our parameter regime, those effects are very small and
    do not change the main results.

 \section{Results and analysis}
  For the understanding of the decay of two interacting bosons, a good starting point is the
  parameter region where we know the exact analytic solutions. In our case, we know exact eigenfunctions
  of Hamiltonian for two extreme cases, $g=0$ and $g=\infty$. In those cases, the two particle eigenfunctions are
  obtained by the combination of one particle eigenfunctions, which are known in analytic form. For arbitrary $g>0$, the results
   lie between these two extremes, and the exact analytic forms are not known.

   \subsection{Vanishing and infinite interaction limits}

  When $g=0$ there is no interaction between two particles. They act independently, with only a symmetric wavefunction condition. The eigenfunction is written as
 \beqa
  \psi (k_1,k_2,x_1,x_2) =\frac{1}{\sqrt 2} ( \phi (k_1,x_1)\phi (k_2,x_2) + \phi (k_2,x_1)\phi (k_1,x_2) ),
  \eeqa
 where the total eigenenergy is $E= (k_1^2 + k_2^2)/2$ and $\phi (k,x)$ is the one particle eigenfunction with eigen wavevector $k$. In our model the explicit form of $\phi$ is given by
  \beqa
  \phi(k, x) =\left\{ \begin{array}{ll} c_1 (k) \sin (k x)
  \;& \mbox{for $0< x<1$} \\
  c_2 (k) e^{i k x} + c_3 (k) e^{-i k x} \; &\mbox{for $1 \le x$ }
   \end{array} \right.
  \eeqa
 where
  \beqa
 & & c_1(k)= \sqrt{ \frac{2}{\pi}} \frac{1}{\sqrt{ 1+ \frac{4 V}{k} \sin k \cos k + \frac{4 V^2}{k^2} \sin^2 k}}, \\
  & &c_2(k) =  \frac{1}{2} \bigg( -(i + \frac{V}{k} ) + \frac{V}{k} e^{-2ik}\bigg) c_1(k), \\
  & & c_3 (k) =  \frac{1}{2}\bigg( (i - \frac{V}{k} ) + \frac{V}{k} e^{2ik}\bigg) c_1(k).
   \eeqa
  The one particle decay rate can be calculated by outgoing boundary condition,  setting the coefficient of outgoing wave $c_3 (k)=0$ and solving for $k$ (this is also the pole of scattering matrix).
  The equation $c_3 (k)=0$ has complex solutions, each of them corresponds to different decay modes.
 We denote the complex solution of  $c_3 (k)=0$ as $k_{z0}, k_{z1},...$ with $k_{z0}$ the lowest decay mode and $k_{z1}$ next lowest decay mode, etc. For the $V=\infty$ ground state initial condition
 \beqa
  \psi_{ini} (x) = \sqrt{2} \sin (\pi x),
  \eeqa
 the dominant decay mode is $k_{z0}$. Since the decay mode wavefunction is also a complex eigenfunction, its time dependence is given by
 $e^{-i E t}$, where $E= k_{z0}^2 /2$. The time evolution of one particle probability inside the potential trap $P_{\rm 1in}(t)$ follows the exponential decay
  \beqa
  & &P_{\rm 1in}(t) \approx | e^{-i E t}|^2 = e^{-\gam_{z0} t}, \\
  & &\gam_{z0} = - 2k_{z0r} k_{z0i},
  \eeqa
 where $k_{z0r}$ and $k_{z0i}$ are the real and imaginary parts of $k_{z0}$, respectively.

     The decay of two interacting bosons in the special cases of $g=0$ and $g=\infty$ is obtained from the single particle decay patterns, respectively.
     
     For $g=0$, the two particle wave function is the product of one-particle wave functions, and their decay is just the product of the individual decay.
     With the condition that the initial wave function was the ground state of $V =\infty$:
  \beqa
   \psi_{\rm ini} (x_1,x_2) =2 \sin (\pi x_1) \sin (\pi x_2).
   \eeqa
If we write the probability of both particle inside the trap as $P_1$, probability of one particle in and one out as $P_2$ and both particles out as $P_3$, their dominant time evolutions are
     \beqa
     & &P_1 (t) \approx e^{-2 \gam_{z0} t}, \EQN{g0P1}, \\
     & &P_2(t)\approx 2e^{-\gam_{z0} t} (1-e^{-\gam_{z0}t}) \EQN{g0P2}, \\
     & &P_3(t)\approx (1- e^{-\gam_{z0} t})^2 \EQN{g0P3}.
     \eeqa

  Another case we know the exact eigenfunction of Hamiltonian is $g=\infty$ case. In this case, the
  two particle eigenfunction is written as
   \beqa
   & &\psi (k_1,k_2,x_1,x_2) =\frac{1}{\sqrt{2}}( \phi (k_1 x_1)\phi(k_2 x_2) -\phi(k_2 x_1) \phi(k_1 x_2) ),
   \nonumber \\
    & &\mbox{for $x_1 \ge x_2$},
   \eeqa
 and $ \psi (k_1,k_2,x_1,x_2) = \psi (k_1,k_2,x_2,x_1)$ for  the $x_1<x_2$ region. Like in the case of fermions the probability density is zero along $x_1=x_2$ line. The $V_0 =\infty$ ground state initial condition is given by
  \beqa
   & &\psi_{\rm ini} (x_1,x_2) = \sqrt{2} (\sin (\pi x_1) \sin (2\pi x_2) - \sin (2\pi x_1)\sin(\pi x_2) ) \nonumber \\
   & &\mbox{for $x_1 \ge x_2$}
   \eeqa
 and $\psi_{\rm ini} (x_2,x_1)= \psi_{\rm ini} (x_1,x_2)$ for $x_1<x_2$.
  The decay of $g=\infty$ two bosons involves two decay mode, with lowest wavevector $k_{z0}$ and next lowest one
  $k_{z1}$. The time evolutions of $P_1$, $P_2$ and $P_3$ are
  \beqa
  & &P_1(t) \approx e^{-(\gam_{z0}+\gam_{z1})t} \\
  & &P_2(t)\approx e^{-\gam_{z0}t} (1- e^{-\gam_{z1}t}) + e^{-\gam_{z1}t} (1- e^{-\gam_{z0}t}) \\
  & &P_3(t)\approx (1- e^{-\gam_{z0}t})(1- e^{-\gam_{z1}t})
  \eeqa
 with
  \beqa
   \gam_{z_j} = -2 k_{zjr} k_{zji}
   \eeqa
   and $k_{zjr}$ and $k_{zji}$ are real and imaginary parts of $k_{zj}$, respectively.

 \subsection{ Arbitrary $g>0$ case}
 For the general case of $0< g< \infty$ exact analytic eigenfunctions are not known. We use the numerical method of
 section~\ref{numerical} to obtain the decay of probabilities. To conduct the simulation,
 first the initial condition was chosen as the ground state of trap potential strength $V=\infty$ limit.

Quite interestingly, the numerical results in this section show that a rather simple model
 can be used to explain interacting boson decay. For the decay of interacting particles, it is expected
 that the number density of particles shows non-exponential decay. When there are more particles inside the trap
 it decays faster, and with less particles the decay is slower. But if we examine the probability $P_{1}$ of both particles inside and the probability $P_2$ of one particle inside and another out  separately, they show quite distinctive features.

  If we plot the logarithm $\ln P_1$
  vs time for various interaction strength $g$,
  the graphs show straight lines, meaning the decay is exponential.
   Furthermore, the decay rate can be obtained by theoretical estimation. Like the decay rate calculation of one-particle case, we can apply the outgoing boundary condition for the wavefunction in region (1).
    Since the probability of both particles escaping simultaneously is very small due to the repulsive interaction,
    it is ignored. Then the outgoing boundary condition from region (1) to region (2) can be written as follows.

\begin{figure}[hbp]
\includegraphics[width=8 cm,height= 4 cm]{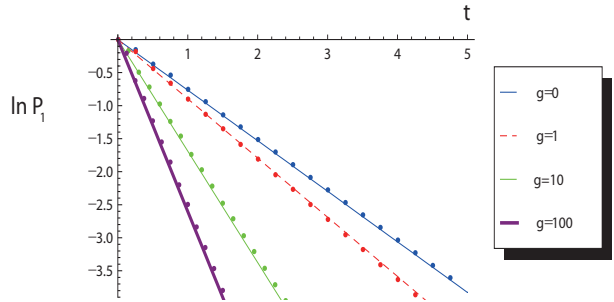}
\caption{$\ln P_1 (t)$ plots for various $g$ (different colors).  Dots are from numerical simulation and lines are
from theoretical prediction of decay rate by outgoing boundary conditions of (\ref{k1g}) and (\ref{k2g}). The
numerical simulation and theoretical prediction show very good agreement. }
\label{LogP1}
\end{figure}

\begin{figure}[hbp]
\includegraphics[width=8 cm,height= 4 cm]{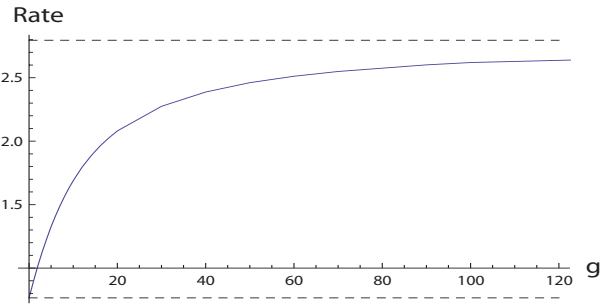}
\caption{$P_1$ decay rate $\gamma_g$ vs $g$ plots. Solid line represents $\gamma_g$, lower and upper dashed lines represent
$P_1$ decay rates of $g=0$ and $g=\infty$ cases, respectively.    }
\label{gammag}
\end{figure}

  First, the wavefunction in region (1)  $\psi_{(1)}$,  satisfying the Bethe ansatz and the boundary conditions at $x_1=0$ and $x_1=x_2$, can be written as [the form of the coefficients without normalization is given in equations (\ref{A1}) to (\ref{A4})]
   \beqa
    & &\psi_{(1)}(x_1,x_2) =  (A_1 (k_1,k_2) e^{i k_{1} x_1} -A_2(k_1,k_2) e^{-i k_{1} x_1}) \sin (k_{2} x_2) \nonumber \\
  & &+ (A_3(k_1,k_2) e^{i k_{2} x_1} -A_4(k_1,k_2) e^{-i k_{2} x_1}) \sin (k_{1} x_2) ,\nonumber \\
   & &\mbox{ for $0 \le x_2 \le x_1 \le 1$}
   \eeqa
 and the outgoing eigenfunction in region (2), $\psi_{(2)}$,  can be written as
  \beqa
  & & \psi_{(2)} (x_1,x_2) =  B_1 e^{i k_1 x_1} \sin (k_2 x_2) + B_2 e^{i k_2 x_1} \sin (k_1 x_2), \nonumber \\
   & & \mbox{for $1< x_1
   $, $0 \le x_2 <1$}
   \eeqa
  with the boundary condition
  \beqa
   & &\psi_{(1)} (1,x_2) = \psi_{(2)} (1,x_2), \EQN{bc12_1} \\
   & &\partial_{x_1} \psi_{(2)} (1,x_2) -  \partial_{x_1} \psi_{(1)} (1,x_2) =2 V \psi_{(1)} (1,x_2). \EQN{bc12_2}
   \eeqa
  Conditions (\ref{bc12_1}) and (\ref{bc12_2}) yield four equations with four unknowns $B_1$, $B_2$, $k_1$ and $k_2$.
  Solving for $k_1$ and $k_2$ we get two equations
  \beqa
 & & A_1(k_1,k_2) e^{ik_1} - A_2(k_1,k_2)e^{-ik_1} =- \frac{i k_1}{V} A_2(k_1,k_2) e^{-i k_1} \EQN{k1g}\\
  & &A_3(k_1,k_2) e^{ik_2} - A_4(k_1,k_2)e^{-ik_2} =- \frac{i k_2}{V} A_4(k_1,k_2) e^{-i k_2}. \EQN{k2g}
  \eeqa
  and two complex wavevectors $k_{1g}$ and $k_{2g}$ for their solutions. When we write real and imaginary parts of complex eigenvectors as
  $k_{1g}= k_{1gr}+ i k_{1gi}$ and $k_{2g}=k_{2gr}+i k_{2gi}$, both of their imaginary parts are negative.
   Considering that the time evolution of an energy eigenfunction follows $e^{-iEt}$
    like the one particle decay mode, it can be expected that $\exp ({- i(k_{1g}^2+ k_{2g}^2)t /2})$
   dominates in time evolution. When we compare the probability of both particle inside $P_1 (t)$
   with $|\exp({- i(k_{1g}^2+ k_{2g}^2)t /2}) |^2 $, indeed we see  that this is what happens.
Both are in very good agreements as shown in figure~\ref{LogP1}. $P_1 (t)$ decays exponentially with
the decay rate predicted by outgoing boundary conditions.
 \beqa
 & & P_1 (t) \approx |\exp({- i(k_{1g}^2+ k_{2g}^2)t /2})|^2 = e^{-\gamma_g t}, \EQN{P1t} \\
  & &\gamma_g = -2k_{1gr}k_{1gi}- 2k_{2gr}k_{2gi}.
  \eeqa
  Figure~\ref{gammag} shows $\gamma_g$ change for various $g$. $\gamma_g$ changes a lot for small $g$, and approaches to
  the decay rate of $g=\infty$ slowly. Figure~\ref{LogP1} shows the comparison between $-\gamma_g t$ line from
  theoretical prediction and $\log P_1$ from numerical simulation.
 They match very well well for all $g>0$, thus showing
 $P_1$ decays exponentially even with interaction between bosons.

 Next we consider the time evolution of $P_2$, one particle in and one particle out of trap probability.
  It is more complicated than that of $P_1$, since it contains probability inflow from region (1) and
  outflow into region (3). Like $P_1$ case, we already know the dominant parts of $P_2 (t)$ for special
  cases, $g=0$ and $g=\infty$.

    For $g=0$, the decay of $P_2(t)$ has the form
    \beqa
    P_{2,g=0} (t) \approx 2 e^{-\gam_{z0} t}(1-e^{-\gam_{z0} t}) = 2 e^{-\gam_{z0} t}- 2 e^{-2 \gam_{z0} t} \EQN{P2g0}
    \eeqa
  and for
  $g=\infty$,
  \beqa
  P_{2,g=\infty} (t) \approx e^{-\gam_{z0} t} (1- e^{-\gam_{z1} t}) +   e^{-\gam_{z1} t} (1- e^{-\gam_{z0} t}).
   \EQN{P2ginf}
   \eeqa
  where $\gam_{z0},\, \gam_{z1}$ are the lowest and next lowest decay rate of one particle in the potential trap.
 For the $g=0$ case, both bosons decay from the same mode independently.
 For the $g=\infty$ case, two bosons decay from the separate decay modes without interfering since they are
 almost orthogonal.
  For general $0< g<\infty$, the time evolution of $P_2(t)$ will be between (\ref{P2g0}) and
  (\ref{P2ginf}) and as $g$ is increased $P_2(t)$ will change from (\ref{P2g0}) to (\ref{P2ginf}).
  We try different models for two regimes where $g$ is not large (weak or moderate repulsion) and where $g$ is very large
  (strong repulsion), and investigate regions of validity for each model.

For the weak or moderate repulsive interaction, we try a simple model for $P_2$ decay.
 If we assume that the probability of both particle escaping simultaneously is very small, which is satisfied
  when the decay rate is small and interparticle interaction is repulsive, then
  the inflow from region (1) has very simple form since the dominant part of $P_1$ satisfies (\ref{P1t}) and
  almost all escaping probability from region (1) goes to region (2).
 We can write $P_2$ as
   \beqa
 \frac{dP_2}{dt} = F_{\rm in} + F_{\rm out}
   \EQN{delP2}
  \eeqa
  where $F_{\rm in}$ is the probability inflow from region (1) to region (2) and $F_{\rm out}$ is the probability
  outflow from region (2) to region (3).

  $F_{\rm in}$ is simply $\gam_g e^{-\gam_g t}$, which is $P_1(t)$ outflow from region (1).
  For the form of outflow $F_{\rm out}$, we try exponential decay model.  In that case, $F_{\rm out}$ is set as $-\gam_{23} P_2$ where $\gam_{23}$ is the decay constant from region (2) to region (3). With this assumption, the solution of (\ref{delP2}) has the form
   \beqa
   P_2 (t) = \frac{\gamma_g}{\gamma_g -\gamma_{23}} (e^{-\gamma_{23} t} - e^{-\gamma_g t}). \EQN{P2onefit}
    \eeqa
   The decay constant $\gam_{23}$ is yet undetermined, so (\ref{P2onefit}) becomes one parameter fitting model.
  This exponential decay model of $F_{out}$  implies that remaining particle in the trap will decay exponentially
    like one particle decay after other one escapes, with only one decay mode.

\begin{figure}[hbp]
\centering
\subfloat[g=0]{\label{figP2g0} \includegraphics[width=.3\textwidth]{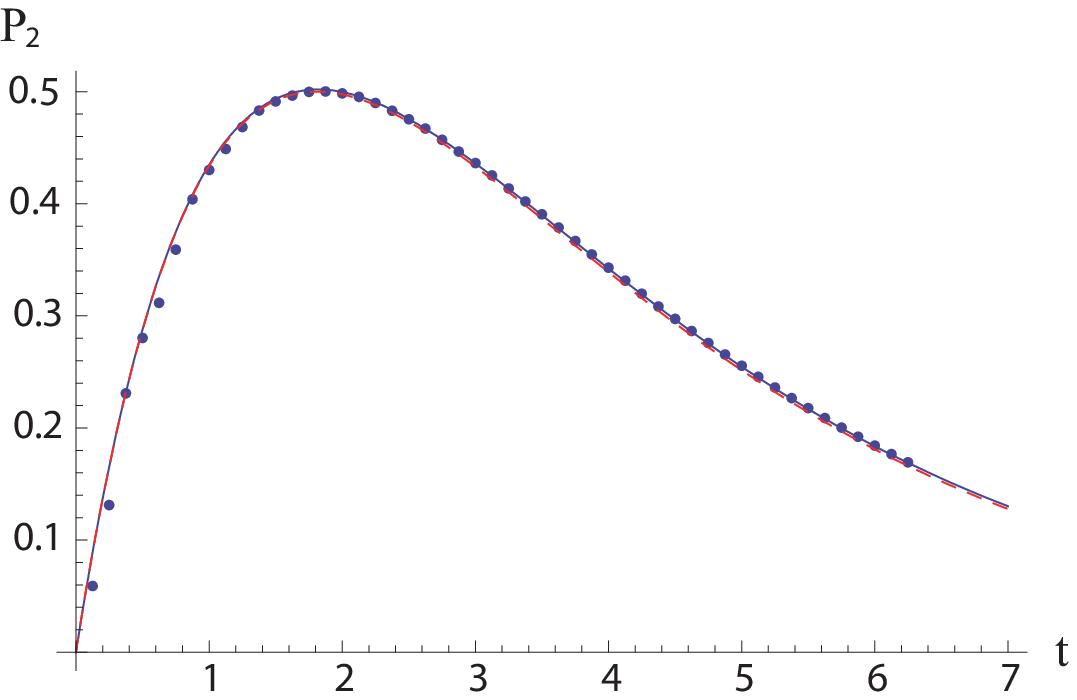}}
\subfloat[g=1]{\label{figP2g2} \includegraphics[width=.3\textwidth]{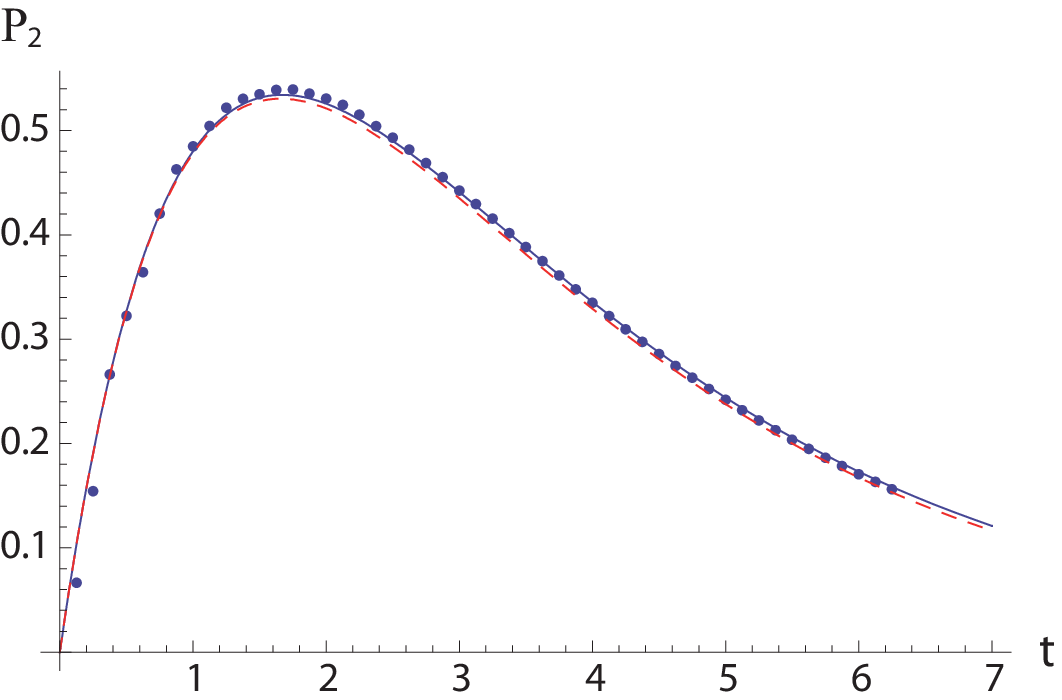}}
\subfloat[g=10]{\label{figP2g10}\includegraphics[width=.3\textwidth]{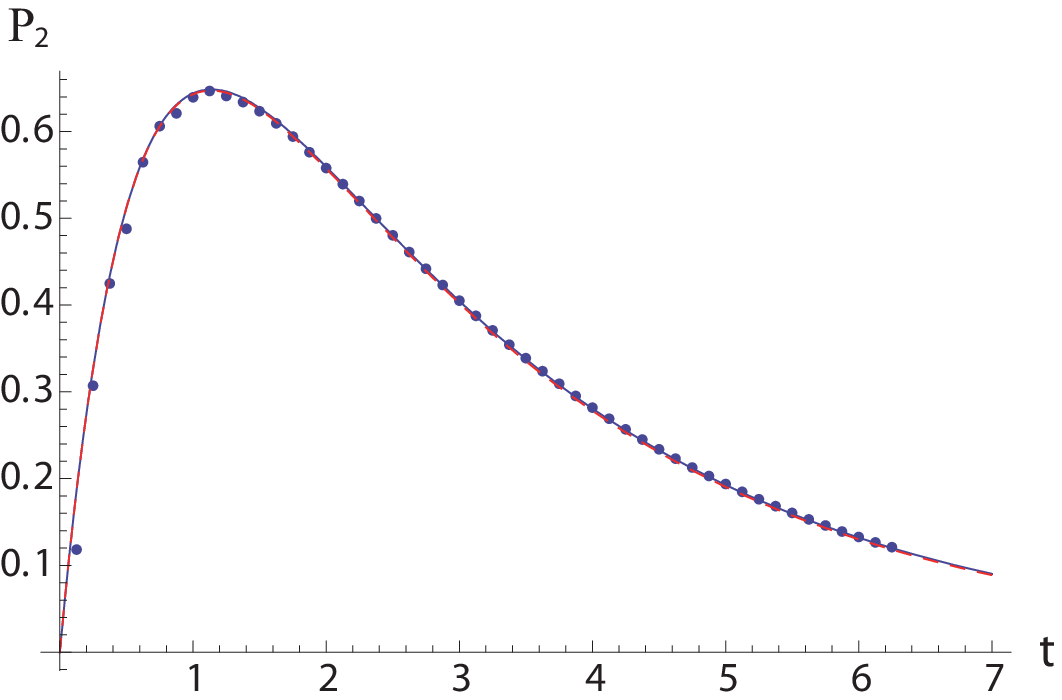} }
\caption{ $P_2 (t)$ plots from (\ref{P2onefit}) (line) and numerical simulation (dots) for $g=0,1,10$.
Dots represent $P_2(t)$ from numerical simulation, dashed red lines are from (\ref{P2onefit}) with $\gam_{23}$ is substituted
by lowest decay rate of single particle and blue lines are from (\ref{P2onefit}) with $\gam_{23}$ obtained
from fitting. All three show very good agreements and lines are almost overlapping. }
\label{P12g0g10}
\end{figure}

\begin{figure}[hbp]
\centering
\subfloat[]{\label{Gam23fit} \includegraphics[width=.4\textwidth]{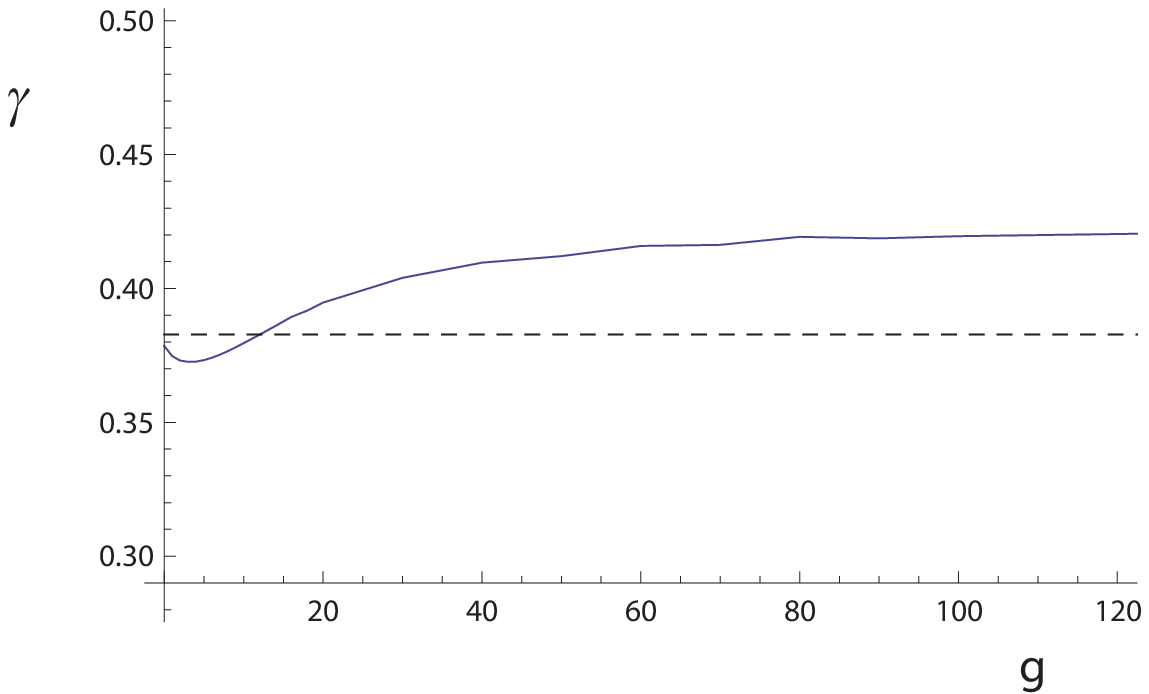}}
\subfloat[]{\label{relGam23Gam10} \includegraphics[width=.4\textwidth]{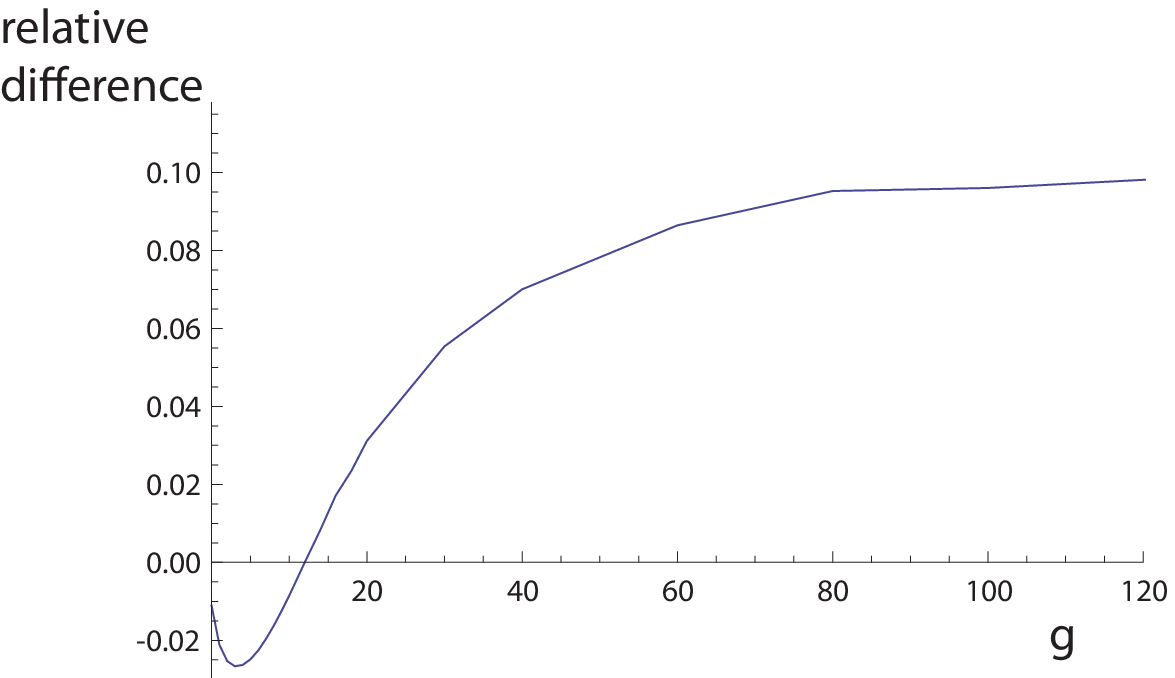}}
\caption{(a) The fitted decay rate $\gam_{23}$ (solid line)
 and the lowest single particle decay rate $\gam_{z0}$ (dashed line) vs $g$. (b) the relative difference
 $(\gam_{23}-\gam_{z0})/ \gam_{z0}$. }
\label{Gam23rel}
\end{figure}

Compared with numerical simulation, model (\ref{P2onefit}) shows very good agreements. Furthermore, it shows that
even for larger $g$ the fitted parameter $\gamma_{23}$ is very close to the lowest
single particle decay rate $\gam_{z0}$.  Figure~\ref{P12g0g10} shows the comparison between numerical
simulation and (\ref{P2onefit}) with $\gamma_{23}$ substituted by $\gam_{z0}$ (dashed red line) and (\ref{P2onefit})
with $\gamma_{23}$ obtained from fitting (blue line), for $g=0,1,10$. All shows very good agreements and
blue lines are not shown well due to overlapping with red. The agreements for even $g=10$ is quite surprising,
since for $g=10$ the initial wavevectors inside trap are far from lowest decay modes as shown in figure~\ref{kgplot}. The initial two wavevector $k_{1i}=5.347$
and $k_{2i}=2.720$, the escaping complex eigenvectors $k_{1g}=4.996-0.1445 i$ and $k_{2g}= 2.507-0.04881 i$
(up to 4 significant digits) for $g=10$. $k_{1i}$ and $k_{1g}$ are closer to second decay modes, but still
$P_2$ decay to region (3) is dominated by single particle lowest decay rate.
Figure~\ref{Gam23rel} shows $\gamma_{23}$ compared to $\gam_{z0}$ and their relative differences for various
$g$. It shows that the relative difference between  $\gamma_{23}$ and $\gam_{z0}$ are less than 2\%
for $0<g<17$, and the difference increases and approaches to 10\% for larger $g$.

 In the strongly repulsive interaction region where the difference between  $\gamma_{23}$ and $\gam_{z0}$ increases,
 the deviation of model \ref{P2onefit} from numerical simulation also increases. In this region we try different model
 which is close to (\ref{P2ginf}).
 Physical meaning of (\ref{P2ginf}) is that there are two decay modes and two decay modes decay independently
 without interfering each other. In our case, we have two complex eigenvector $k_{1g}$ and
 $k_{2g}$ from (\ref{k1g}) and
 (\ref{k2g}). Assuming that $P_2$ decays from each complex wavevector and each mode do not interfere each other,
  we write decay model of $P_2 (t)$
 for large $g$ as
  \beqa
  P_2 (t) =  e^{-\gam_{1g} t}(1-e^{-\gam_{2g}t}) + e^{-\gam_{2g} t}(1-e^{-\gam_{1g}t}). \EQN{P2gbig}
   \eeqa
   where
   \beqa
   \gam_{1g} = -2k_{1gr}k_{1gi},\; \gam_{2g}=-2k_{2gr}k_{2gi} \EQN{gam1g2g}
   \eeqa

\begin{figure}[hbp]
\centering
\subfloat[g=60]{\label{g60fitTwo} \includegraphics[width=.4\textwidth]{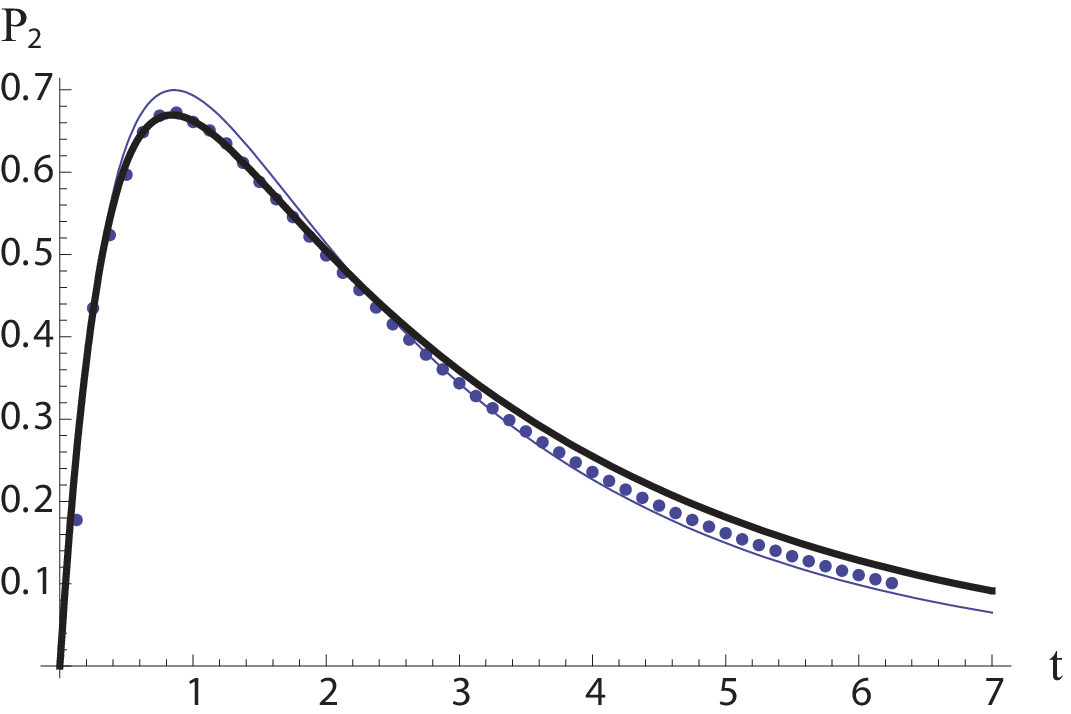}}
\subfloat[g=100]{\label{g100fitTwo} \includegraphics[width=.4\textwidth]{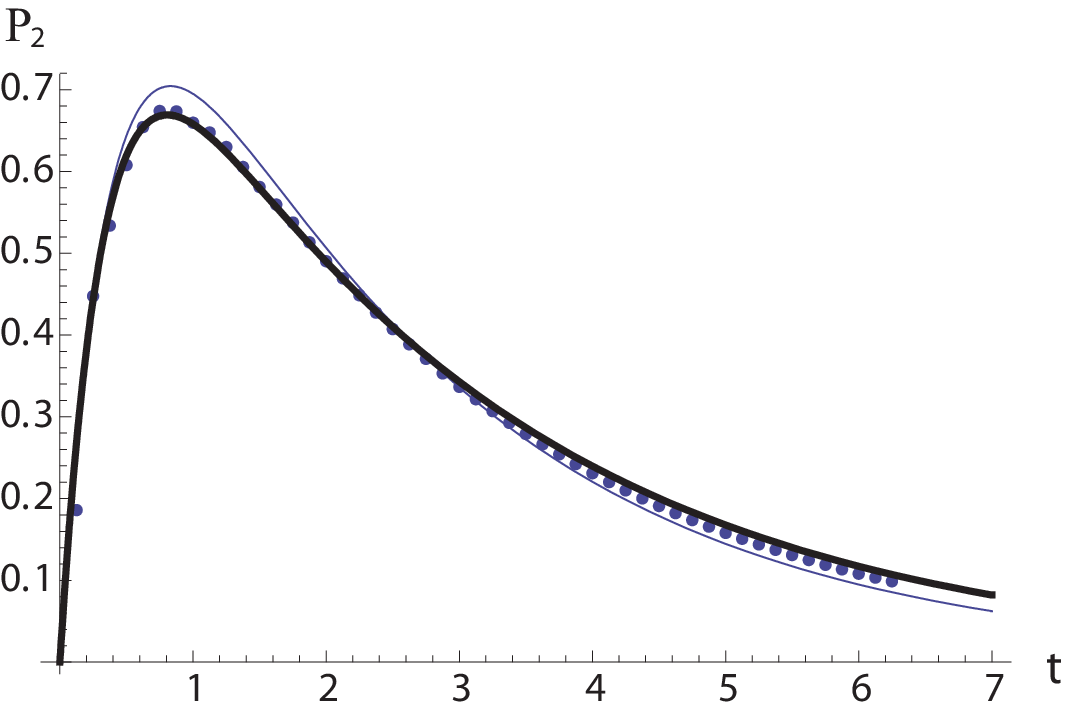}}
\caption{Comparison between numerical simulation (dots) and theoretical models.
Black line represents (\ref{P2gbig}) model and blue line represents (\ref{P2onefit}) model. (\ref{P2gbig}) model
shows better agreement with numerical simulation for larger $g$.
  }
\label{twomodel}
\end{figure}

 This model works better for larger $g$ than  (\ref{P2onefit}) as figure~\ref{twomodel} shows.
 The (\ref{P2gbig}) model describes the peak of $P_2(t)$ well, and discrepancy with the numerical data becomes smaller for larger $g$.

   To see  the agreements between numerical simulation and fitting model quantitatively, we consider the absolute mean of relative
    error $\eta $ between
   the probability calculated by numerical simulation $P_{\rm 2 num}$ and the probability calculated by
    model $P_{\rm 2model}$, defined as
    \beqa
    \eta= \frac{1}{N} \sum_{i}^N \frac{| P_{\rm 2num}(t_i)-P_{\rm 2model} (t_i) |}{P_{\rm 2num}(t_i)} \EQN{error}
    \eeqa
 with $t_i$s are taken from $t=0.1$ to $t=5$ by $0.1$ intervals.

\begin{figure}
        \includegraphics[scale=.7]{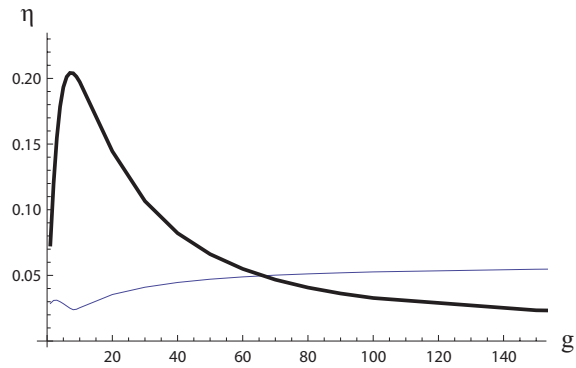}
   \caption{Absolute mean of relative error $\eta$ plots of model (\ref{P2onefit}) (blue) and model (\ref{P2gbig}) (thick black) versus $g$.}
   \label{figP2trelerr2}
   \end{figure}

  Figure~\ref{figP2trelerr2} shows plots of $\eta$ versus interaction strength $g$ plots for
two different models.
 Model (\ref{P2onefit}) shows good agreements with simulation up to $g=17$, with relative error less than 3\%.
 The error increases steadily, reaching more than 5\% when $g>60$ (seen from figure~\ref{kgplot}, this is strongly
 repulsive region).
 Second model (\ref{P2gbig}) shows large deviation from the numerical simulation for smaller $g$, but agreements
 with the simulation becomes better than that of (\ref{P2onefit}) model for $g>67$.

  Finally,  The probability of both particles outside, $P_3$, is easily calculated since
    $P_1+P_2+P_3=1$. So total decay mechanism can be described by (\ref{P1t}), (\ref{P2onefit}) or (\ref{P2gbig}).

\begin{figure}
        \includegraphics[scale=.7]{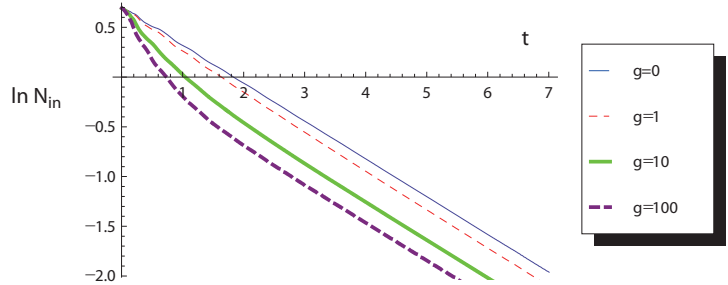}
   \caption{$\ln N_{\rm in} (t)$ plots. For smaller $g$ it is closer to the straight line (exponential decay)
   but for larger $g$ the decay rate changes from faster to slower ones. }
   \label{figLogNt}
   \end{figure}

    If we calculate the number density $N_{\rm in}(t)$ inside the potential, it is written as
    \beqa
    N_{\rm in} (t) = \int_{\rm in} dx \int dx_1 dx_2 \psi^* (x_1,x_2,t) \sum_{i=1}^2 \delta(x-x_i) \psi (x_1,x_2,t)
    \eeqa
 and in our case it simply becomes $N_{\rm in} (t) = 2 P_1(t) + P_2(t)$. Figure~\ref{figLogNt} Shows the logarithm of
 $N_{\rm in}$ versus time. For $g=0,1$ the decay of $N_{\rm in}(t)$ is close to exponential
 ($\ln N_{\rm in}$ close to straight line) but for larger $g$
 it is more visible that the decay rate changes from faster to slower ones, as expected.

\section{Conclusion}
We have calculated the decay of  two repulsively interacting bosons, initially in the ground state of a potential trap,
 by numerical simulation. We have found an exponential decay mode for the probability of both bosons inside the
    trap and have estimated its decay rate theoretically.
    By applying outgoing boundary condition for the loss of single particle from the trap, we obtain two complex wavevectors corresponding to the two particles inside the trap and corresponding decay rate.
    The agreement between numerical simulation and theoretical estimation in time evolution of decay
    probabilities is very good.
For describing the probability to have a single particle inside and one particle outside, two simple models were
     proposed.
     For small and moderate $g$, we apply a model in which the remaining particle decays exponentially, whereas
for larger $g$ (strongly repulsive) we propose another model in which the modes of each complex wave vector decay separately.
Our numerical simulations show very good agreement for weak and moderate interactions with the first model. For stronger interactions, where fermionization effects become relevant,
separate exponential decay model becomes necessary and agrees well with simulations.
  The number density shows
       that the decay rate changes over time from fast to slower decay for large $g$.
 The results show that  simple models describe the overall decay mechanism of repulsively
 interacting bosons well.

\acknowledgements
 Authors thank to A. Dudarev for helpful discussions.   JB was supported by the Marsden Fund Council (Contract No.\ MAU0706) from Government funding administered by the Royal Society of New Zealand.

\end{document}